\documentclass[12pt,a4paper,UTF8]{article}
\usepackage{amsmath,mathtools,amsfonts,amsmath,amssymb,mathrsfs,dsfont,cases,amsthm,bm,bbm}
\usepackage{color,xcolor}
\usepackage{graphicx}
\usepackage{titlesec}
\usepackage{geometry}
\usepackage[colorlinks,citecolor=blue,urlcolor=blue,linkcolor=blue]{hyperref}%
\usepackage{array,booktabs,makecell,tabularx,multirow}  

\newcommand{\cX}{\mathcal{X}}
\newcommand{\cY}{\mathcal{Y}}
\newcommand{\bbP}{\mathbb{P}}
\newcommand{\bbE}{\mathbb{E}}
\newcommand{\cN}{\mathcal{N}}

\newtheoremstyle{thry}
{1em}
{1em}
{\itshape\rmfamily}
{}
{\scshape\large}
{.}
{.5em}
{}
\theoremstyle{thry}

\newtheorem{theorem}{\indent Theorem}
\newtheorem{lemma}{\indent  Lemma}
\newtheorem{assumption}{\indent  Assumption}
\newtheorem{corollary}{\indent  Corollary}

\title{Two--Sample Test for Stochastic Block Models via Maximum Entry--Wise Deviation}
\author{Kang Fu, Seydou Keita and Jianwei Hu \\ {\small \itshape School of Mathematics and Statistics, Central China Normal University, Wuhan}}
\date{\today}

\begin{document}
\maketitle
\begin{abstract}
The stochastic block model is a popular tool for detecting community structures in network data. Detecting the difference between two community structures is an important issue for stochastic block models. However, the two-sample test has been a largely under-explored domain, and too little work has been devoted to it. In this article, based on the maximum entry--wise deviation of the two centered and rescaled adjacency matrices, we propose a novel test statistic to test two samples of stochastic block models. We prove that the null distribution of the proposed test statistic converges in distribution to a Gumbel distribution, and we show the change of the two samples from stochastic block models can be tested via the proposed method. Then, we show that the proposed test has an asymptotic power guarantee against alternative models. One noticeable advantage of the proposed test statistic is that the number of communities can be allowed to grow linearly up to a logarithmic factor. Further, we extend the proposed method to the degree-corrected stochastic block model. Both simulation studies and real-world data examples indicate that the proposed method works well.

\noindent
\textbf{Keywords:} Gumbel distribution; Network data; Stochastic block model; Two-sample test
\end{abstract}

\section{Introduction}

Network data analysis has become a popular research topic in many fields, including gene classification, social relationship investigation, and financial risk management. In the past decades, the majority of works mainly focused on the large-scale network data with community structure, see, e.g., Newman \& Girvan (2004); Newman (2006); Steinhaeuser \& Chawla (2010). In the network data analysis, the stochastic block model proposed by Holland et al. (1983) is a popular tool to fit the network data with community structure, see, e.g., Snijders \& Nowicki (1997); Nowicki \& Snijders (2001); Bickel \& Chen (2009); Rohe et al. (2011); Choi et al. (2012); Jin (2015); Zhang \& Zhou (2016). The stochastic block model with $K$ communities assumes that $n$ nodes of the network are clustered into $K$ communities, that is, there exists a mapping of community membership (also known as community membership label) $g: [n]\rightarrow [K]^n$,  where $[n] = \{1,\ldots,n\}$. Formally, $g(i) = k$ means that the node $i$ belongs to the community $k$. For an unweighted and undirected graph $G$, it can be represented by a binary symmetric adjacency matrix $A$, that is, $A_{ij}=1$ if there is a connection (or an edge) between node $i$ and node $j$ and $A_{ij}=0$ otherwise. Given the community membership label $g$, the stochastic block model assumes that the entries $A_{ij}(i>j)$ of the adjacency matrix $A$ are mutually independent Bernoulli random variables with probabilities $P_{ij} = B_{g(i)g(j)}$ for a certain symmetric probability matrix $B \in [0, 1]^{K\times K}$, where matrix $P$ is called as edge-probability matrix. Then the stochastic block model is completely and uniquely determined by the pair $(g, B)$ up to label permutations of nodes.

The fundamental issues in the stochastic block model are model selection and community detection. Given an adjacency matrix $A$, the goal of model selection is to estimate the number of clusters or communities, and the goal of community detection is to cluster all nodes into different communities such that the connections between the nodes in the same community are dense and the connections between the nodes in the different communities are sparse. To enhance the flexibility of the model, variations of the stochastic block model were also proposed, such as the degree-corrected stochastic block model proposed by Karrer \& Newman (2011) addressed the degree heterogeneity by introducing the additional node activeness parameters, and Airoldi et al. (2008) proposed the mixed membership stochastic block model, where a single node may belong to multiple communities. For the model selection, many methods are used to estimate the number of communities, including the sequential testing methods (Lei, 2016; Hu et al., 2021), and the likelihood-based methods (Sald\~{n}a et al., 2017; Wang \& Bickel, 2017; Hu et al., 2020). Meanwhile, the majority of efficient methods also have been proposed to recover the community structure, such as modularity (Newman, 2006), variational methods (Daudin et al., 2008), profile-likelihood maximization (Bickel \& Chen, 2009), spectral clustering (Rohe et al., 2011; Jin, 2015), pseudo-likelihood maximization (Amini et al., 2013), and profile-pseudo likelihood methods (Wang et al., 2021). The corresponding asymptotic properties of estimation of community label have been obtained, see, e.g., Rohe et al. (2011); Zhao et al. (2012); Choi et al. (2012); Lei \& Rinaldo (2015); Sarkar \& Bickel (2015); Zhang \& Zhou (2016); Wang et al. (2021).

In the past decades, the research interest of much literature focuses on the study of one sample of stochastic block models. Usually, in the social network, people surrounding a leader may tend to develop a closer relationship with another leader,  see, e.g., Barnett \& Onnela (2016). This phenomenon may lead to an issue, that is, whether the community structure (the number of communities $K$, the community probability matrix $B$, and the community membership label $g$) of the network will change over time or the environment.

As the simplest case of the multi-layer stochastic block model, two sample networks may also appear. For a multi-layer stochastic block model, it can be classified into a variety of more specific situations: First, all layer networks come from the identical stochastic block model $(g,B)$; second, each layer of the network comes from a different model but they have the same community structure, that is, $g$ is identical across all layers; third, each layer of the network comes from different models, that is, $g$ and $B$ are different in each layer. In fact, for the third case, it can be considered as a mixed model of multiple stochastic block models. However, how could we distinguish the three situations? Intuitively, we should judge whether two models are the same when we have two samples from the corresponding models. A common inference method to judge whether two models are the same is the hypothesis testing method. Tang et al. (2017) considered whether two independent finite-dimensional random dot product graphs are generated by the identical model. Using the adjacency spectral embedding for each sample, they constructed a statistic based on the kernel function. Ghoshdastidar \& von Luxburg (2018) used the largest singular value of a scaled and centralized matrix to construct the statistic and proved that the null distribution convergences in distribution to a Tracy-Widom distribution. Ghoshdastidar et al. (2020) proposed two test statistics using the Frobenius norm and spectral norm to test whether two samples of networks are generated from the identical edge-probability matrix. However, this testing procedure requires choosing an appropriate threshold. Recently, Chen et al. (2021a) simplified the statistics in Ghoshdastidar \& von Luxburg (2018) and proposed a test procedure for a two-sample network test. Based on the random matrix theory, Chen et al. (2021b) used the trace of a constructed matrix to obtain the statistic and proved that the asymptotic null distribution is the standard normal distribution.

For the methods mentioned above, one defect is that either the edge-probability matrices of the two populations differ greatly or multiple samples are required. Hence, when we only have two samples, the methods mentioned above may not work well as there is less information. In addition, it is worth noting that when the community structure changes slightly, the edge-probability matrix may change less, especially when the network size is relatively large. The above methods of testing $H_0: P_1=P_2$ may not work well, where $P_1$ and $P_2$ are the edge-probability matrices of two network models. It implies that we cannot test the difference between the two models when the difference is tiny but the network size is very large. Thus, detecting the difference between two samples of stochastic block models is an interesting research problem. Under the two-sample test of the stochastic block model, Wu et al. (2022) proposed a test method based on the locally smoothed adjacency matrix. To construct the smoothed adjacency matrix, their method separates a community into serval non-overlapping neighboring sub-communities and averages the entries of the adjacency matrix in non-overlapping local neighborhoods within communities. However, the procedure is complex and only applicable to a small number of communities. In this article, we want to construct a statistic that allows $K$ to diverge with $n$.

In this article, based on the maximum entry-wise deviation of the two centered and rescaled adjacency matrices, we propose a two-sample test statistic to detect the change of two stochastic block models under two observed adjacency matrices. We show that the asymptotic null distribution of the test statistic is a Gumbel distribution when $K = o(n/\log^2 n)$. This testing method allows $K$ to grow linearly with $n$ up to a logarithmic factor. It is well known that the number of communities must be less than the number of nodes in the stochastic block model. Since $K$ cannot grow faster, Rohe et al. (2014) called this scenario ($K\rightarrow\infty$) the highest dimensional stochastic block model. Compared with the method in Wu et al.(2022), we relax the condition that the number of communities $K$ is fixed. Moreover, we also show that the proposed test is asymptotically powerful against serval alternative models, and does not need additional methods to improve the power of the proposed test. Finally, to improve the flexibility of the proposed method, we extend the proposed method to the degree-corrected stochastic block model. 

Next, we formally describe this issue. Let $\cX=(X_{ij})_{n\times n}$ and $\cY=(Y_{ij})_{n\times n}$ be two binary symmetric adjacency matrices from two stochastic block models parametrized by $(g_x, B_x)$ and $(g_y, B_y)$, respectively. For any $i$, let $X_{ii} = Y_{ii} = 0$, i.e., no loop edge exists. In this article, we assume that the node label of the two networks is identical, instead of the community label of the node. Hence, the two networks $\cX$ and $\cY$ can be viewed as repeated observations in the same individuals. Then, given two sample networks $\cX$ and $\cY$, the testing problem can be formulated as
\begin{equation}\label{eq:problem}
	H_0: (g_x, B_x) = (g_y, B_y)\qquad \mathrm{v.s.}\qquad H_1: (g_x, B_x) \neq (g_y, B_y).
\end{equation}
Intuitively, there are four mutually exclusive scenarios for the alternative hypothesis: (i) $K_x \neq K_y$, (ii) $K_x = K_y, g_x = g_y$, but $B_x \neq B_y$, (iii) $K_x = K_y, B_x = B_y$, but $g_x \neq g_y$, and (iv) $K_x = K_y, g_x\neq g_y, B_x\neq B_y$. 

Our testing procedure for \eqref{eq:problem} goes as follows. First, we get consistent estimators of $K_x$ and $K_y$, denoted by $\widehat{K}_x$ and $\widehat{K}_y$, respectively. If $\widehat{K}_x\neq\widehat{K}_y$, it means that $(g_{x},B_{x})=(g_{y},B_{y})$ cannot be true, so we can directly reject the null hypothesis. If $\widehat{K}_x=\widehat{K}_y$, based on $\widehat{K}_x=\widehat{K}_y=\widehat{K}$, we obtain the strongly consistent estimators $\widehat{g}_x$ and $\widehat{g}_y$ for the community membership labels $g_x$ and $g_y$. Second, we estimate the entries of $B_x$ and $B_y$ by the sample proportions of each community based on $(\cX, \widehat{g}_x)$ and $(\cY, \widehat{g}_y)$. Third, we use $\widehat{B}_{y}$ ($\widehat{B}_{x}$) to center and rescale adjacency matrix $\cX$ ($\cY$), and sum for new matrix under specific rules, and get a combined information matrix, see \eqref{CIM}. Finally, we obtain the test statistic by the maximum of the elements of the combined information matrix. The basic principle of this method is that if the null hypothesis is true, the entries of the combined information matrix asymptotically follow the normal distribution. According to the results of Zhou (2007), the asymptotic distribution of the maximum of elements after normalization is a Gumbel distribution. On the other hand, under the alternative hypothesis, the adjacency is incorrectly centered and scaled, and this deviation is magnified to a very large value by the normalization term. This implies that the test statistic can successfully separate $H_0$ and $H_1$ in \eqref{eq:problem}.

The remainder of the article is organized as follows. In Section \ref{method}, we introduce the new test statistic and state its asymptotic null distribution and asymptotic power. In Section \ref{Extend}, we extend the proposed method to the degree-corrected stochastic block model. Simulation studies and real-world data examples are given in Sections \ref{Simulation} and \ref{Real}, respectively. All technical proofs are postponed to the Appendix.

\section{A NEW TWO-SAMPLE TEST FOR THE STOCHASTIC BLOCK MODEL}\label{method}
Consider a stochastic block model on $n$ nodes with the community label $g$ and probability matrix $B$. Given a number of communities $K$ and a community label $g$, the maximum likelihood estimator of $B$ is given by
\begin{equation}\label{eq:estB}
	\widehat{B}_{uv} = \begin{dcases}
 \dfrac{\sum_{i\in\cN_{u}, j\in\cN_{v}}X_{ij}}{n_u n_v}, \quad & u \neq v,\\
 \dfrac{\sum_{i,j\in\cN_{u},i\neq j}X_{ij}}{n_u (n_v-1)}, \quad & u = v,	
 \end{dcases}
\end{equation}
where $\cN_{u} = \left\{i: g(i) = u\right\}$ for $1 \leq u \leq K$, $i\in\{1,\ldots, n\}$ stands for the label of a node, and $n_u = \left|\cN_{u}\right|$. After here, $\mathcal{N}_u^x=\left\{i:g_x(i) = u\right\}$ and $n_u^x = \left|\cN^x_{u}\right|$, where $x$ can be replaced by $y$.

For an adjacency matrix $A$, Hu et al. (2021) used a test statistic, based on the maximum entry-wise deviation, to test the following two hypothesis tests:

(1) $H_0:K = K_0\quad \mathrm{v.s.}\quad H_1: K> K_0$, and 

(2) $H_0:g = g_0\quad \mathrm{v.s.}\quad H_1: g\neq g_0$,

\noindent
where $K$ and $g$ denote the true number of communities and the true community label, respectively, and $K_0$ and $g_0$ denote a hypothetical number of communities and hypothetical community label, respectively. Let
\[
L_n(K_0, g_0) := \max_{1\leq i\leq n,1\leq v\leq K_0}|\widehat{\rho}_{iv}|,
\]
where $\widehat{\rho}_{iv} = \dfrac{1}{\sqrt{|g^{-1}_0(v)/\{i\}|}}\sum_{j\in g^{-1}_0(v)/\{i\}}\dfrac{A_{ij}-\widehat{B}_{g_0(i)g_0(j)}}{\sqrt{\widehat{B}_{g_0(i)g_0(j)}(1-\widehat{B}_{g_0(i)g_0(j)})}}$, and $g_0^{-1}(v) = \{i:g_0(i)=v\}$, and $\left|g_0^{-1}(v)\right|$ is the number of nodes in block $v$, and $g_0^{-1}(v)/\{i\}$ denotes the set of nodes that belong to community $v$ in $g_x$ but excluding node $i$ and $\widehat{B}_{g_0(i)g_0(j)}$ as defined in \eqref{eq:estB}. Under the null hypothesis $H_0: K = K_0, g = g_0$, if $K = o(n/\log^2 n)$, Hu et al. (2021) showed that
\[
\lim_{n\rightarrow\infty}\bbP\{L_n^2(K_0, g_0) - 2\log(2K_0 n) +\log\log(2K_0n)\leq y\} = \exp\left\{-\dfrac{1}{2\sqrt{\pi}}e^{-y/2}\right\},
\]
and proposed the following test statistic:
\[
T_n = L_n^2(K_0, g_0) - 2\log(2K_0 n) +\log\log(2K_0n).
\]
Then the corresponding level-$\alpha$ rejection rule is
\[
\mathrm{Reject:}\ H_0: K = K_0, g = g_0\ \mathrm{if}\ T_n\geq t_{(1-\alpha)},
\]
where $Q_{1-\alpha}$ is the $\alpha$th quantile of the Gumbel distribution with $\mu = -2\log(2\sqrt{\pi})$ and $\beta = 2$.

In this article, We aim to develop a new test statistic that allows it can be used to test two samples. For two samples $\cX$ and $\cY$ from stochastic block models $(g_x, B_x)$ and $(g_y, B_y)$, respectively, let $\widehat{K}_x$ and $\widehat{K}_y$ be obtained by some estimation methods, such as the recursive approach in Zhao et al. (2011), the sequential testing method given in Lei (2016), the likelihood-based method in Sald\~{n}a et al. (2017), the corrected Bayesian information criterion in Hu et al. (2020), the test based on maximum entry-wise deviation in Hu et al. (2021), and the spectral methods in Le \& Levina (2022). In this article, we use the corrected Bayesian information criterion in Hu et al. (2020) to get consistent estimators $\widehat{K}_x$ and $\widehat{K}_y$. Recall that given the number of communities $K_x$ and $K_y$, the community labels $g_x$ and $g_y$ can be consistently estimated, denote by $\widehat{g}_x$ and $\widehat{g}_y$, by some existing strongly consistent community detection procedures, e.g., the majority voting algorithm in Gao et al. (2017) and the profile-pseudo likelihood method in Wang et al. (2021). In view of this, throughout the paper, we formally assume that 
\begin{equation}\label{eq:consistent}
\bbP\left\{\widehat{g}_x = g_x, \widehat{g}_y = g_y, \widehat{K}_x = K_x, \widehat{K}_y = K_y\right\}\rightarrow1,
\end{equation}
which implies that we require that all estimators are strongly consistent. Hence, we can get the $\widehat{B}_x$ and $\widehat{B}_y$ by equation \eqref{eq:estB}. Note that the community membership label $\widehat{g}_x$ and the probability matrix $\widehat{B}_x$ depend on the number of communities $\widehat{K}_x$. If $\widehat{g}_x$ is a strongly consistent estimator, $\widehat{K}_x$ must be the consistent estimator. Then, Condition \eqref{eq:consistent} can be relaxed to $\bbP\left\{\widehat{g}_x = g_x, \widehat{g}_y = g_y\right\}\rightarrow1$. In fact, if $\widehat{K}_x\neq\widehat{K}_y$, then it is natural that $\widehat{g}_x \neq \widehat{g}_y$ and $\widehat{B}_x\neq\widehat{B}_y$. Since $\widehat{g}_x$ and $ \widehat{g}_y$ are consistent estimators, we can reject the null hypothesis with probability tending to 1. Meanwhile, For case (iv), there is indeed a problem of identifiability. Because both $g$ and $B$ are allowed to vary, the existing methods cannot reasonably solve the problem of identifiability. Thus, in this article, we mainly focus on cases (ii) and (iii) for four alternative hypotheses. Then, we assume $K_x = K_y = K$. Without loss of generality, we write $\widehat{K} = \widehat{K}_x = \widehat{K}_y$ when $\widehat{K}_x = \widehat{K}_y$. Next, we formally state the test statistic. As mentioned in the introduction, our test statistic is motivated by the contrast of $\cX$ (or $\cY$) and $\widehat{B}_y$ (or $\widehat{B}_x$), i.e.:

\begin{multline}\label{CIM}
	\tilde{\rho}_{iv} (g_x, g_y) = \dfrac{1}{\sqrt{\left|g_x^{-1}(v)/\{i\}\right|+\left|g_y^{-1}(v)/\{i\}\right|}}\times \\ \left[\sum\limits_{j\in g_y^{-1}(v)/\{i\}}\dfrac{X_{ij} - \widehat{B}^y_{g_y(i) g_y(j)}}{\sqrt{\widehat{B}^y_{g_y(i)g_y(j)}\left(1-\widehat{B}^y_{g_y(i)g_y(j)}\right)}}+\sum\limits_{j\in g_x^{-1}(v)/\{i\}}\dfrac{Y_{ij} - \widehat{B}^x_{g_x(i)g_x(j)}}{\sqrt{\widehat{B}^x_{g_x(i)g_x(j)}\left(1-\widehat{B}^x_{g_x(i)g_x(j)}\right)}}\right]
\end{multline} 

Moreover, based on the maximum entry-wise deviation, the proposed test statistic has the following form:
\[
L_n(g_x, g_y) = \max_{1\leq i\leq n, 1\leq v\leq K}\left|\tilde{\rho}_{iv}(g_x, g_y)\right|.
\]

\subsection{The Asymptotic Null Distribution}

To obtain the asymptotic result for the statistic $L_n(g_x, g_y)$, we first make the following assumptions:

\begin{assumption}
	The entries of $B_x$ and $B_y$ are uniformly bounded away from 0 and 1, and both $B_x$ and $B_y$ have no identical rows.
\end{assumption}

\begin{assumption}
	There exist $C_1$, $C_2$, $C_3$ and $C_4$ such that
	\[
	C_1n/K_x\leq\min_{1\leq u\leq K_x}n_u^x\leq\max_{1\leq u\leq K_x}n_u^x\leq C_2n^2/(K_x^2\log^2n),
	\]
	and,
	\[
	C_3n/K_y\leq\min_{1\leq u\leq K_y}n_u^y\leq\max_{1\leq u\leq K_y}n_u^y\leq C_4n^2/(K_y^2\log^2n),
	\]
	for all $n$.
\end{assumption}

Assumption 1 requires that the entries of the probability matrices $B_x$ and $B_y$ are uniformly bounded away from 0 and 1. This assumption is similar to the corresponding condition in Lei (2016) and Hu et al. (2021). Meantime, Assumption 1 also requires that $B_x$ and $B_y$ are identifiable, which is a basic condition (Wang \& Bickel, 2017). Assumption 2 not only requires a lower bound for the number of nodes in the smallest community, but also gives an upper bound for the number of nodes in the largest community. The lower bound requires that the number of nodes in the smallest community for $\cX$ ($\cY$) is at least proportional to $n/K_x$ ($n/K_y$). This is a mild assumption and easy to be achieved. For example, this lower bound can be achieved when the community label $g_x$ is generated from a multinomial distribution with $n$ trials and parameter $\bm{\pi} = (\pi_1, \cdots, \pi_K)$ such that $\min_{u}\pi_u\geq C_1/K_x$. For the assumption about the upper bound, Zhang \& Zhou (2016) and Gao et al. (2017) also considered a similar condition, which is used to control the maximum within-group deviation between the $B_x$ ($B_y$) and its estimator $\widehat{B}_x$ ($\widehat{B}_y$).

We now give the asymptotic property of the test statistic $L_n(g_x, g_y)$ and delay the proof to the Appendix.

\begin{theorem}\label{Thm:gumbel}
	Suppose that Assumptions (1) and (2) hold. Then under the null hypothesis $H_0: (g_x, B_x) = (g_y, B_y)$, as $n\rightarrow\infty$, if $K = o(n/\log^2n)$, we have 
	\begin{equation}\label{eq:gumbel}
	\bbP\left\{L_n^2(g_x, g_y)-2\log(2Kn)+\log\log(2Kn)\leq y\right\}\rightarrow\exp\left\{-\dfrac{1}{2\sqrt{\pi}}e^{-y/2}\right\},
\end{equation}
where the right hand-side of \eqref{eq:gumbel} is the cumulative distribution function of the Gumbel distribution with $\mu = -2\log(2\sqrt{\pi})$ and $\beta = 2$.
\end{theorem}

Note that Theorem 1 demonstrates that $L_n^2(g_x,g_y) - 2\log(2Kn) + \log\log(2Kn)$ converges in distribution to a Gumbel distribution under $H_0$. 

Using the above Theorem 1, we can implement hypothesis test \eqref{eq:problem} as follows. First, we estimate the number of communities $\widehat{K}_x$ and $\widehat{K}_y$, and the community membership labels $(\widehat{g}_x, \widehat{B}_x)$ and $(\widehat{g}_y, \widehat{B}_y)$, respectively. Then, we compute the statistic
\[
T_n = L_n^2(\widehat{g}_x, \widehat{g}_y)-2\log(2\widehat{K}n)+\log\log(2\widehat{K}n).
\]
Since $\widehat{g}_x$ and $\widehat{g}_y$ are strongly consistent estimators, we have that $T_n$ intuitively follows the Gumbel distribution with $\mu = -2\log(2\sqrt{\pi})$ and $\beta = 2$. To carry out the hypothesis test, we have a rejection rule:
\[
	\mathrm{Reject}\ \ H_0, \quad T_n\geq Q_{1-\alpha},
\]
where $Q_{1-\alpha}$ is the $\alpha$th quantile of the Gumbel distribution with $\mu = -2\log(2\sqrt{\pi})$ and $\beta = 2$. In Section \ref{Simulation}, by simulation studies, we investigate the finite sample performance of the proposed test statistic.

In general, the null distribution converges to the Gumbel distribution slowly, which is also confirmed in simulation studies. Bickel \& Sarkar (2015) considered a bootstrap method for the correction of the distribution of the finite sample, which was also used by Lei (2016) and Hu et al. (2021). For our statistic, the bootstrap corrected test statistic is calculated as follows:

1. Using the consistent estimation method to estimate $\widehat{K}_x = \widehat{K}_y = \widehat{K}$, then get the estimation $(\widehat{g}_x,\widehat{B}_x)$ and $(\widehat{g}_y,\widehat{B}_y)$ by the strongly consistent clustering method;

2. For $m = 1,\ldots,M$, generate $\cX^{(m)}$ and $\cY^{(m)}$ from the edge-probability matrix $\widehat{P}_{ij} = (\widehat{P}_{ij}^x + \widehat{P}_{ij}^y)/2 = (\widehat{B}_{\widehat{g}_x(i)\widehat{g}_x(j)}^x + \widehat{B}_{\widehat{g}_y(i)\widehat{g}_y(j)}^y)/2$, and calculate $T_n^{(m)}$ based on $\cX^{(m)}$, $\cY^{(m)}$;

3. Using $\{T_n^{(m)}: m=1,\ldots,M\}$ to estimate the location and scale parameters $\widehat{\mu}$ and $\widehat{\beta}$ of the Gumbel distribution through maximum likelihood method.

4. The bootstrap corrected test statistic is calculated as
\[
T_{n}^{boot} = \mu + \beta\left(\dfrac{T_n - \widehat{\mu}}{\widehat{\beta}}\right),
\]
where $\mu = -2\log(2\sqrt{\pi})$ and $\beta = 2$.

\subsection{THE ASYMPTOTIC POWER}
In this subsection, we investigate the asymptotic power of the proposed test procedure. To guarantee good testing power, we need the following theoretical assumption.
\begin{assumption}
	The maximum grouped difference between $B_x$ and $B_y$ satisfies:
	\[
	\max_{i,v}|\dfrac{1}{\sqrt{|g_x^{-1}(v)|}}\sum_{j\in g_x^{-1}(v)}(B_{g_y(i)g_y(j)}^y-B_{g_x(i)g_x(j)}^x)|/\sqrt{\log n}\rightarrow\infty,
	\]
	and
	\[
	\max_{i,v}|\dfrac{1}{\sqrt{|g_y^{-1}(v)|}}\sum_{j\in g_y^{-1}(v)}(B_{g_x(i)g_x(j)}^x-B_{g_y(i)g_y(j)}^y)|/\sqrt{\log n}\rightarrow\infty.
	\]
\end{assumption}

Assumption 3 specifies that under the alternative $H_1$, the maximum grouped difference between $B_x$ and $B_y$ diverges faster than $\sqrt{\log n}$. This assumption is analogous to (A2$'$) in Hu et al. (2021). Then, the lower bound of the growth rate of the test statistic $T_n$ under the alternative hypothesis $H_1: (g_x, B_x)\neq(g_y, B_y)$ is given in the following Theorem.

\begin{theorem}
	Suppose that Assumptions (1),  (2), and (3) hold. Then under the alternative hypothesis $H_1: (g_x, B_x) \neq (g_y, B_y)$, as $n\rightarrow\infty$, if $K=o(n/\log^2n)$, we have 
\[
L_n^2(g_x,g_y)\geq 2\log(2Kn)-\log\log(2Kn) + c_1\log n + O_p(1),
\]
for some some consistent $c_1 > 0$.
\end{theorem}

The proof is collected in the Appendix. The above Theorem 2 shows that $T_n$ diverges faster than $\log n$ as $n\rightarrow\infty$ under the alternative hypothesis $H_1$. Then, we have the following corollary saying that the asymptotic size is close to the nominal level and the asymptotic power is almost equal to 1.

\begin{corollary}
	Suppose that Assumptions (1), and (2) hold, given a nominal level $\alpha$, we have
	\[
	\bbP_{H_0}\{T_n > Q_{1-\alpha}\}\rightarrow\alpha,
	\]
	and further, suppose that Assumption (3) holds, we have
	\[
	\bbP_{H_1}\{T_n > Q_{1-\alpha}\}\rightarrow1.
	\]
\end{corollary}

Therefore, the asymptotic null distribution in Theorem 1 and the growth rate in Theorem 2 suggest that the null hypothesis and the alternative hypothesis are well separated, and our proposed test is asymptotically powerful against alternative hypothesis $H_1: (g_x, B_x) \neq (g_y, B_y)$.

\section{EXTENSION}\label{Extend}
In this section, we extend the proposed method to the degree-corrected stochastic block model. In reality, a network may contain many high-degree nodes and lower-degree nodes, that is, the degree heterogeneity of nodes. To address this issue, Karrer \& Newman (2011) proposed a degree-corrected stochastic block model. Specifically, for an undirected network $\cX$, this model assume that the edge $X_{ij}$ satisfies $\bbP\{X_{ij}=1\}=\theta_{i}\theta_{j}B_{g(i)g(j)}$, where $\theta_{i}$ is the degree parameter for node $i$. To ensure the identifiability of the model, we assume that $\sum_{i}\theta_i/n=1$. For two sample networks $\cX$ and $\cY$, the testing problem has the following form:
\begin{equation}\label{eq:DCtestproblem}
	H'_0: (g_x, B_x,\theta_x) = (g_y, B_y,\theta_y)\qquad \mathrm{v.s.}\qquad H'_1: (g_x, B_x,\theta_x) \neq (g_y, B_y,\theta_y),
\end{equation}
where $\theta_x$ and $\theta_y$ are the node degree parameters for networks $\cX$ and $\cY$, respectively.

Similar to the general stochastic block model, for the two-sample test of the degree-corrected stochastic block model, we consider the following test statistic:
\[
L_{D}(g_x,g_y) = \max_{i,v}|\tau_{iv}|,
\]
where 
\begin{multline}\label{CIM}
	\tau_{iv} = \dfrac{1}{\sqrt{\left|g_x^{-1}(v)/\{i\}\right|+\left|g_y^{-1}(v)/\{i\}\right|}}\times \\ \left[\sum\limits_{j\in g_y^{-1}(v)/\{i\}}\dfrac{X_{ij} - \widehat{\theta}^y_{i}\widehat{\theta}^y_{j}\widehat{B}^y_{g_y(i) g_y(j)}}{\sqrt{\widehat{\theta}^y_{i}\widehat{\theta}^y_{j}\widehat{B}^y_{g_y(i)g_y(j)}\left(1-\widehat{\theta}^y_{i}\widehat{\theta}^y_{j}\widehat{B}^y_{g_y(i)g_y(j)}\right)}}+\right.\\
	\left.\sum\limits_{j\in g_x^{-1}(v)/\{i\}}\dfrac{Y_{ij} - \widehat{\theta}^x_{i}\widehat{\theta}^x_{j}\widehat{B}^x_{g_x(i)g_x(j)}}{\sqrt{\widehat{\theta}^x_{i}\widehat{\theta}^x_{j}\widehat{B}^x_{g_x(i)g_x(j)}\left(1-\widehat{\theta}^x_{i}\widehat{\theta}^x_{j}\widehat{B}^x_{g_x(i)g_x(j)}\right)}}\right],
\end{multline}
where $\widehat{\theta}^x_i=|g_x^{-1}(u)|d_i/\sum_{j\in\{j:g_x(j)=g_x(i)\}}d_j$ is the maximum likelihood estimator of $\theta^x_i$ and $d_i=\sum_{j}X_{ij}$. $\widehat{\theta}^y_i$ is similar to that of $\widehat{\theta}^x$. Note that it is difficult to obtain the asymptotic null distribution of $L_D(g_x,g_y)$ as the complex dependency between the entries $\tau_{iv}$. By the simulation studies, we find that the empirical distribution of $T_{D}=L_D^2(\widehat{g}_x, \widehat{g}_y)-2\log(2\widehat{K}n)+\log\log(2\widehat{K}n)$ is also the Gumbel distribution with $\mu = -2\log(2\sqrt{\pi})$ and $\beta = 2$. Then, we have a rejection rule:
\[
	\mathrm{Reject}\ \ H'_0, \quad T_D\geq Q_{1-\alpha},
\]
where $Q_{1-\alpha}$ is the $\alpha$th quantile of the Gumbel distribution with $\mu = -2\log(2\sqrt{\pi})$ and $\beta = 2$. Similar to the general stochastic block model, we can obtain the bootstrap corrected statistic through the identical method.

\section{SIMULATION}\label{Simulation}

In this section, we evaluate the performance of the proposed test statistics in various simulation studies. Firstly, the number of communities is estimated by the Bayesian information criterion proposed by Hu et al. (2020), and the community membership label is estimated by strongly consistent estimation methods mentioned above, i.e., the profile-pseudo likelihood method in Wang et al. (2022), initialized by spectral clustering with permutations, is used to obtain the community membership label. In the simulation, we consider the test statistic $T_n = L_n^2(\widehat{g}_x, \widehat{g}_y)-2\log(2\widehat{K}n)+\log\log(2\widehat{K}n)$ and the bootstrap corrected test statistic $T_n^{boot}$. In comparative experiments, the maximum deviation method (referred to as TST-MD) proposed by Chen et al. (2021a) and the spectral-based method (referred to as TST-S) proposed by Chen et al. (2021b) are used.

\subsection{Simulation 1: The Null Distribution Under the Stochastic Block Model}

In the first simulation, we consider the finite sample null distribution of test statistic $T_n$ and empirically verify Theorem 1 for two samples from a common stochastic block model. Since the limiting distribution of the test statistic is proven to be a Gumbel distribution, the location and scale parameters can be estimated by generating some bootstrap samples to correct the test statistic at a lower cost. In all of our simulations, we use $M = 100$. 

\begin{figure}[h]
	\includegraphics[width = 8cm]{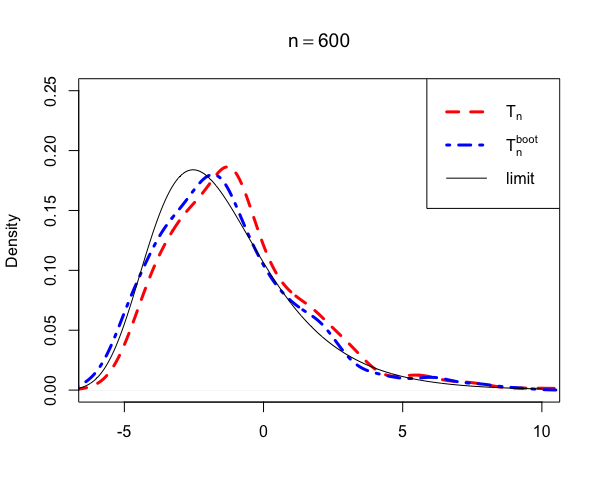}
	\includegraphics[width = 8cm]{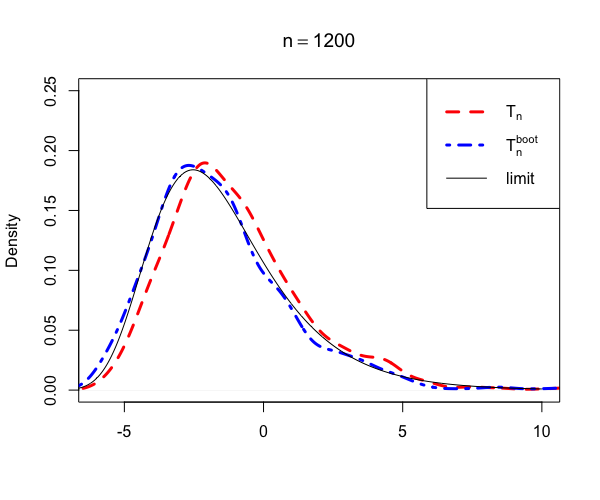}
	\caption{Null densities under the stochastic block model in Simulation 1 with $n = 600$ (left plot) and $n = 1200$ (right plot). The red dashed lines, blue dash-dotted lines, and black solid lines show the densities of the test statistic $T_n$, the bootstrap corrected test statistic $T_n^{boot}$, and the theoretical limit, respectively.}
	\label{figure:gumbel}
\end{figure}

We generate data from the stochastic block model with $B_{uv}^x = B_{uv}^y = 0.1 (1 + 4\times\mathds{1}(u=v))$. In this setting, we set $K_x = K_y = 3$ with $\pi_1 = \pi_2 = \pi_3 = 1/3$. The sample size is either $n = 600$ or $n = 1200$. We use the strongly consistent method to get $\widehat{g}_x$ and $\widehat{g}_y$, then get $\widehat{B}_x$ and $\widehat{B}_y$, e.g., the profile-pseudo likelihood method in Wang et al. (2021). In Figure \ref{figure:gumbel}, based on 1000 data replications, we plot the sample distribution of the test statistic $T_n$ with and without bootstrap correction. Figure \ref{figure:gumbel} demonstrates that the empirical probability density function of $T_n$ biases upward, but the bias decrease as the sample size increases. Compared with the distribution of $T_n$, the distribution of $T_n^{boot}$ fits the true distribution better.

\subsection{Simulation 2: Empirical Size for Hypothesis \eqref{eq:problem}}

In this subsection, we study the empirical size under varying $K_x$, $K_y$, $B_x$ and $B_y$. We set the edge probability between communities $u$ and $v$ as $B_{uv}^x = B_{uv}^y = r (1 + 3\times\mathds{1}(u=v))$, where $r$ measures the sparsity of the network. Let $K_x, K_y \in \{2,3,4,5,6,10,15,20,30\}, r\in\{0.05, 0.1, 0.2\}$, and the size of each block be 200. Table \ref{Table:size} reports the result from 200 data replications. From Table \ref{Table:size}, $T_n$'s Type I errors are close to the nominal level when $K$ is small, $T_n^{boot}$'s Type I errors are close to the nominal level even when $K$ is large. Compared with TST-MD and TST-S, the performances of $T_n^{boot}$ and TST-S are better than TST-MD. In some cases, TST-MS does not work well. Inversely, the empirical size of $T_n^{boot}$ and TST-S are close to the nominal level. It is worth noting that as the sample size increases, the empirical size of the $T_n$ gradually increases, which seems to be contrary to fundamental theory. In fact, as the sample size increases, the number of communities also increases. As a result, the commonly used community label estimation algorithms cannot exactly recover the community label. But by the bootstrap correction, the bootstrap corrected statistic $T_n^{boot}$ has a good empirical size, which implies that the statistic with the bootstrap correction works well. We also notice that when the network is sparse, the test statistic $T_n$ does not have good empirical size, and tends to be a little oversized.

\begin{table}[h]
\setlength{\belowcaptionskip}{0.5cm}
\centering
\caption{Empirical size at nominal level $\alpha = 0.05$ for hypothesis test $H_0: (g_x, B_x) = (g_y, B_y)\ \mathrm{v.s.}\ H_1: (g_x, B_x) \neq (g_y, B_y)$. In the results of $T_n^{boot}$, the values in the parentheses are the empirical size of TST-MD (left) and TST-S (right), respectively.}
\label{Table:size}
\begin{tabular}{cccccccc}
\hline
\multirow{2}{*}{} & \multicolumn{3}{c}{$T_n$} & & \multicolumn{3}{c}{$T_n^{boot}$}    \\ \cline{2-4} \cline{6-8} 
& $r=0.05$ & $r=0.1$ & $r=0.2$ & & $r=0.05$ & $r=0.1$ & $r=0.2$ \\ \hline
$K=2$ & 0.04 & 0.09 & 0.03 & & 0.04 (0.09,0.03) & 0.08 (0.1,0.02) & 0.05 (0.01,0.05) \\
$K=3$ & 0.09 & 0.06 & 0.07 & & 0.05 (0.085,0.04) & 0.05 (0.04,0.03) & 0.04 (0.01,0.03) \\
$K=4$ & 0.09 & 0.08 & 0.06 & & 0.04 (0.01,0.04) & 0.05 (0,0.03) & 0.05 (0,0.04) \\
$K=5$ & 0.09 & 0.09 & 0.05 & & 0.04 (0,0.04) & 0.06 (0,0.04) & 0.06 (0,0.03) \\
$K=6$ & 0.08 & 0.09 & 0.04 & & 0.04 (0.005,0.05) & 0.04 (0,0.05) & 0.04 (0,0.03) \\
$K=10$ & 0.20 & 0.10 & 0.06 & & 0.06 (0,0.04) & 0.05 (0,0.04) & 0.04 (0,0.06) \\
$K=15$ & 0.46 & 0.16 & 0.09 & & 0.08 (0.01,0.04) & 0.05 (0,0.05) & 0.05 (0,0.04) \\
$K=20$ & 0.77 & 0.14 & 0.10 & & 0.10 (0.03,0.04) & 0.02 (0,0.05) & 0.08 (0,0.05) \\
$K=30$ & 0.99 & 0.23 & 0.13 & & 0.05 (0.04,0.05) & 0.02 (0.005,0.06) & 0.04 (0,0.05) \\ 
\hline
\end{tabular}
\end{table}

\subsection{Simulation 3: Empirical Power for Hypothesis \eqref{eq:problem}}

In this subsection, we investigate the empirical power under hypothesis test \eqref{eq:problem}. We consider two kinds of alternatives: i) $B_x \neq B_y$ but $g_x = g_y$ and ii) $g_x \neq g_y$ but $B_x = B_y$. Similar to simulation 2, let $K_x, K_y\in\{2,3,4,5,6,10,20,30\}$. The sample size is either $n = 600$ or $n = 1200$. For the first alternative, we generate $\cX$ with edge possibility $3.5r$ within community and $0.5r$ between communities, and $\cY$ with edge possibility $8r$ within community and $3r$ between communities for $r\in\{0.01, 0.05, 0.1\}$. For the second alternative, we generate $\cX$ and $\cY$ with edge probability $3r$ within community and $r$ between communities for $r\in\{0.05, 0.1, 0.2\}$. We set $g_x = (1, \ldots, 1, \ldots, K_x, \ldots, K_x)$ and $g_y$ from the multinomial distribution with $n$ trials and probability $\bm{\pi} = (1/K_y, \cdots, 1/K_y)$. Similarly, all simulations are run 200 times. For both alternatives, Tables \ref{table:power1} - \ref{table:power2} report the empirical power for hypothesis test \eqref{eq:problem}. We observe that the proposed test statistics and TST-MD can successfully detect all alternative hypotheses. In the second alternative hypothesis, however, the empirical power of TST-S is less than 1. As discussed in the introduction, if the change of community structure is small, there will be little difference between the two edge-probability matrices. As a result, the TST-S will not separate the null hypothesis and the alternative hypothesis well.

\begin{table}[h]
\setlength{\belowcaptionskip}{0.5cm}
\centering
\caption{Empirical power of $T_n$ at nominal level $\alpha = 0.05$ for the first alternative. The values in the parentheses are the empirical power of $T_n^{boot}$(left), TST-MD (middle), and TST-S (right), respectively.}
\label{table:power1}
\begin{tabular}{ccccccc}
\hline
\multirow{2}{*}{} & \multicolumn{3}{c}{$n=600$}   & \multicolumn{3}{c}{$n=1200$}  \\ \cline{2-7} 
                  & $r=0.01$ & $r=0.05$ & $r=0.1$ & $r=0.01$ & $r=0.05$ & $r=0.1$ \\ \hline
$K=2$ & 1 (1,1,1) & 1 (1,1,1) & 1 (1,1,1) & 1 (1,1,1) & 1 (1,1,1) & 1 (1,1,1) \\
$K=3$ & 1 (1,1,1) & 1 (1,1,1) & 1 (1,1,1) & 1 (1,1,1) & 1 (1,1,1) & 1 (1,1,1) \\
$K=4$ & 1 (1,1,1) & 1 (1,1,1) & 1 (1,1,1) & 1 (1,1,1) & 1 (1,1,1) & 1 (1,1,1) \\
$K=5$ & 1 (1,1,1) & 1 (1,1,1) & 1 (1,1,1) & 1 (1,1,1) & 1 (1,1,1) & 1 (1,1,1) \\
$K=6$ & 1 (1,1,1) & 1 (1,1,1) & 1 (1,1,1) & 1 (1,1,1) & 1 (1,1,1) & 1 (1,1,1) \\
$K=10$ & 1 (1,1,1) & 1 (1,1,1) & 1 (1,1,1) & 1 (1,1,1) & 1 (1,1,1) & 1 (1,1,1) \\
$K=15$ & 1 (1,1,1) & 1 (1,1,1) & 1 (1,1,1) & 1 (1,1,1) & 1 (1,1,1) & 1 (1,1,1) \\
$K=20$ & 1 (1,1,1) & 1 (1,1,1) & 1 (1,1,1) & 1 (1,1,1) & 1 (1,1,1) & 1 (1,1,1) \\
$K=30$ & 1 (1,1,1) & 1 (1,1,1) & 1 (1,1,1) & 1 (1,1,1) & 1 (1,1,1) & 1 (1,1,1) \\ \hline
\end{tabular}
\end{table}

\begin{table}[h]
\setlength{\belowcaptionskip}{0.5cm}
\centering
\caption{Empirical power of $T_n$ at nominal level $\alpha = 0.05$ for the second alternative. The values in the parentheses are the empirical power of $T_n^{boot}$(left), TST-MD (middle), and TST-S (right), respectively.}
\label{table:power2}
\begin{tabular}{ccccccc}
\hline
\multirow{2}{*}{} & \multicolumn{3}{c}{$n=600$}   & \multicolumn{3}{c}{$n=1200$}  \\ \cline{2-7} 
                  & $r=0.01$ & $r=0.05$ & $r=0.1$ & $r=0.01$ & $r=0.05$ & $r=0.1$ \\ \hline
$K=2$ & 1 (1,1,0.04) & 1 (1,1,0.17) & 1 (1,1,0.54) & 1 (1,1,0.04) & 1 (1,1,0.27) & 1 (1,1,0.57) \\
$K=3$ & 1 (1,1,0.06) & 1 (1,1,0.21) & 1 (1,1,0.51) & 1 (1,1,0.02) & 1 (1,1,0.39) & 1 (1,1,0.55) \\
$K=4$ & 1 (1,1,0.045) & 1 (1,1,0.16) & 1 (1,1,0.45) & 1 (1,1,0.09) & 1 (1,1,0.35) & 1 (1,1,0.63) \\
$K=5$ & 1 (1,1,0.09) & 1 (1,1,0.13) & 1 (1,1,0.32) & 1 (1,1,0.07) & 1 (1,1,0.33) & 1 (1,1,0.58) \\
$K=6$ & 1 (1,1,0.06) & 1 (1,1,0.08) & 1 (1,1,0.19) & 1 (1,1,0.05) & 1 (1,1,018) & 1 (1,1,0.5) \\
$K=10$ & 1 (1,0.91,0.09) & 1 (1,1,0.06) & 1 (1,1,0.09) & 1 (1,1,0.08) & 1 (1,1,0.07) & 1 (1,1,0.09) \\
$K=15$ & 1 (1,1,0.07) & 1 (1,0.65,0.07) & 1 (1,0.99,0.09) & 1 (1,1,0.08) & 1 (1,1,0.07) & 1 (1,1,0.1) \\
$K=20$ & 1 (1,1,0.12) & 1 (1,0.82,0.06) & 1 (1,0.79,0.06) & 1 (1,1,0.09) & 1 (1,1,0.07) & 1 (1,1,0.1) \\
$K=30$ & 1 (1,1,0.1) & 1 (1,0.95,0.09) & 1 (1,0.75,0.07) & 1 (1,1,0.73) & 1 (1,1,0.06) & 1 (1,1,0.09) \\ \hline
\end{tabular}
\end{table}

\subsection{Simulation 4: The Null Distribution Under the Degree-Corrected Stochastic Block Model}

In this subsection, we investigate the asymptotic null distribution of the proposed statistic $T_D$. Similar to Simulation 1, we also consider the bootstrap corrected statistic. The bootstrap corrected method is similar to the general stochastic block model and is given in the Appendix.

\begin{figure}[h]
	\includegraphics[width = 8cm]{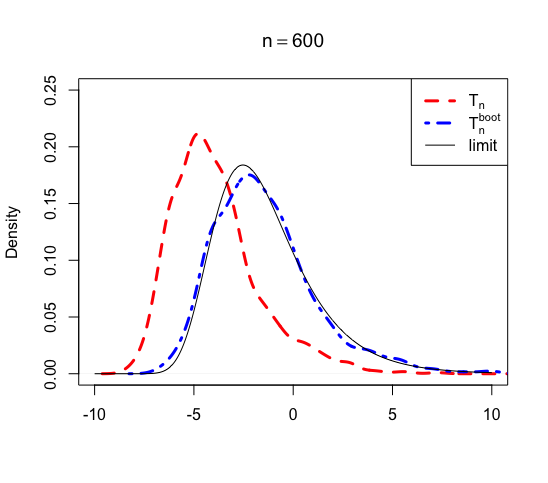}
	\includegraphics[width = 8cm]{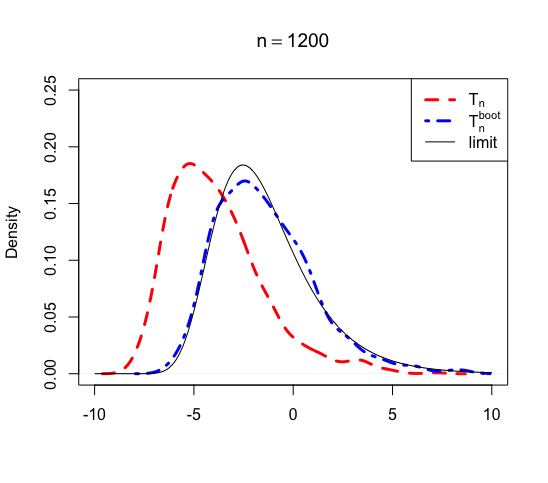}
	\caption{Null densities under the degree-corrected stochastic block model in Simulation 4 with $n = 600$ (left plot) and $n = 1200$ (right plot). The red dashed lines, blue dash-dotted lines, and black solid lines show the densities of the test statistic $T_D$, the bootstrap corrected test statistic $T_D^{boot}$, and the theoretical limit, respectively.}
	\label{figure:gumbelDC}
\end{figure}

As shown in Zhao et al. (2012), we use the same approach to generate the degree corrected parameters $\theta$. The details are as follows:
$$\theta_i=\begin{dcases}\xi_i, & \mathrm{w.p.} 0.8,\\ 9/11, & \mathrm{w.p.} 0.1,\\ 13/11, & \mathrm{w.p.} 0.1,\end{dcases}$$
where $\xi_i$ is a uniformly random variable on the interval $[4/5,6/5]$. Similar to Simulation 1, we set $K_x=K_y=3$ with $\pi_1=\pi_2=\pi_3=1/3$ and $B_{uv}^x = B_{uv}^y = 0.1 (1 + 4\times\mathds{1}(u=v))$. We also consider sample sizes $n=600$ and $1200$. Based on 1000 data replications, we report the results in Figure \ref{figure:gumbelDC}. Figure \ref{figure:gumbelDC} shows the empirical distribution of the test statistic $T_D$ is the Gumbel distribution by a location and scale shift. With the sample size increasing, the deviation between the empirical distribution of $T_D$ and the limiting distribution does not decrease. After using the bootstrap correction, the empirical distribution of $T_D^{boot}$ is close to the limiting distribution. It also implies that the empirical size of the test statistic $T_D$ is smaller than the nominal level.

\subsection{Simulation 5: Empirical Size for Hypothesis \eqref{eq:DCtestproblem}}

In this subsection, we consider the empirical size under the framework of the degree-corrected stochastic block model. The basic settings are similar to Simulation 2 except $K_x,K_y\in\{2,3,4,6\}$. The degree corrected parameters are generated by the method in Simulation 4. Table \ref{Table:sizeDC} shows the simulation results. It is observed that the probability of Type I error is close to the nominal level. At the same time, the empirical size of the statistic $T_D$ is less than that of $T_D^{boot}$, which is also consistent with the results of Simulation 4.
\begin{table}[h]
\setlength{\belowcaptionskip}{0.5cm}
\centering
\caption{Empirical size at nominal level $\alpha = 0.05$ for hypothesis test $H_0: (g_x, B_x, \theta_x) = (g_y, B_y, \theta_y)\ \mathrm{v.s.}\ H_1: (g_x, B_x, \theta_x) \neq (g_y, B_y, \theta_y)$. The values in the parentheses are the empirical power of $T_n^{boot}$.}
\label{Table:sizeDC}
\begin{tabular}{cccccccc}
\hline
\multirow{2}{*}{} & \multicolumn{3}{c}{$T_D$} & & \multicolumn{3}{c}{$T_D^{boot}$}    \\ \cline{2-4} \cline{6-8} 
& $r=0.05$ & $r=0.1$ & $r=0.2$ & & $r=0.05$ & $r=0.1$ & $r=0.2$ \\ \hline
$K=2$ & 0.01 & 0.07 & 0.05 & & 0.04 & 0.05 & 0.06 \\
$K=3$ & 0.01 & 0.01 & 0.04 & & 0.04 & 0.06 & 0.04 \\
$K=4$ & 0.04 & 0.01 & 0.04 & & 0.05 & 0.04 & 0.05 \\
$K=6$ & 0.06 & 0.01 & 0.05 & & 0.06 & 0.04 & 0.05 \\
\hline
\end{tabular}
\end{table}

\subsection{Simulation 6: Empirical Power for Hypothesis \eqref{eq:DCtestproblem}}

In this simulation, we investigate the testing power for the two-sample test under the degree-corrected stochastic block model. We only consider the case of $g_x\neq g_y$. The probability matrix and the community label are generated the same as in Simulation 3. We consider the settings that $n=600$ and $1200$. The degree corrected parameters are similar to Simulation 4. The empirical results are shown in Table \ref{Table:powerDC}. From Table \ref{Table:powerDC}, we can know that the proposed test statistic can successfully detect the alternative hypotheses as the test has good power.

\begin{table}[h]
\setlength{\belowcaptionskip}{0.5cm}
\centering
\caption{Empirical power of $T_D$ at nominal level $\alpha = 0.05$ for hypothesis test $H_0: (g_x, B_x, \theta_x) = (g_y, B_y, \theta_y)\ \mathrm{v.s.}\ H_1: (g_x, B_x, \theta_x) \neq (g_y, B_y, \theta_y)$.}
\label{Table:powerDC}
\begin{tabular}{cccccccc}
\hline
\multirow{2}{*}{} & \multicolumn{3}{c}{$T_D$} & & \multicolumn{3}{c}{$T_D^{boot}$}    \\ \cline{2-4} \cline{6-8} 
& $r=0.05$ & $r=0.1$ & $r=0.2$ & & $r=0.05$ & $r=0.1$ & $r=0.2$ \\ \hline
$K=2$ & 1 & 1 & 1 & & 1 & 1 & 1 \\
$K=3$ & 1 & 1 & 1 & & 1 & 1 & 1 \\
$K=4$ & 1 & 1 & 1 & & 1 & 1 & 1 \\
$K=6$ & 0.98 & 1 & 1 & & 1 & 1 & 1 \\
\hline
\end{tabular}
\end{table}

\section{DATA EXAMPLE}\label{Real}
\subsection{Gene co-expression data}
In this subsection, we apply the proposed method to the gene dataset of developing rhesus monkeys' tissue from the medial prefrontal cortex. This dataset was originally collected by Bakken et al. (2016). Other work analyzing the data suggests this is an appropriate dataset, as other work already has good evidence that gene co-expression patterns in monkey tissues in this brain region change significantly as they develop. For the prenatal period, a 6-layer network is considered, which corresponds to the 6 age stage. We label the 6-layer network as E40 to E120 to indicate the number of embryonic days of age. For the postnatal period, a 5-layer network is considered, which indicates the 5 layers within the medial prefrontal cortex. We label the 5-layer network as L2 to L6. With this data, we aim to show whether two gene co-expression networks in two different periods can be considered to come from a common stochastic block model using the proposed method.

\textbf{Preprocessing Procedure.} The microarray dataset contains $n = 9173$ genes measured among many samples across the $L = 11$ layer. In this article, since we consider the difference between the two samples, we only choose the gene co-expression networks in two periods E40 and E50, that is, the development occurs from 40 days to 50 days in the embryo. Similar to other work, e.g., Langfelder \& Horvath (2008), we preprocess the adjacency matrix as follows. First, to calculate the adjacency matrix, for a layer $l$, we construct the Pearson correlation matrix. Define the co-expression similarity $s_{ij}$ as the correlation coefficient between the profiles of nodes $i$ and $j$: $s_{ij} = cov(x_i, x_j), 1\leq i,j\leq n$. In order to avoid the outliers, let $s_{ij}' = (1 + s_{ij}) / 2$. Then, using a thresholding procedure, the co-expression similarity matrix $S = (s'_{ij})_{n\times n}$ is transformed into the adjacency matrix $\cX$. An unweighted network adjacency $X_{ij}$ between gene expression profiles $x_i$ and $x_j$ can be defined by hard thresholding the co-expression similarity $s'_{ij}$ as
\[
X_{ij} = \begin{dcases} 1 & |s'_{ij}| \geq \tau, \\ 0 & |s'_{ij}| < \tau, \end{dcases}\quad \mathrm{and}\quad X_{ii} = 0,
\]
where $\tau$ is the ``hard" threshold parameter. Thus, two genes are linked ($X_{ij} = 1$) if the absolute correlation between their expression profiles exceeds the hard threshold $\tau$. In this article, we set $\tau = 0.72$ to get two adjacency matrices $\cX$ and $\cY$. Lastly, we remove all the genes corresponding to nodes whose total degree for $\cX$ and $\cY$ is less than 90. The purpose of this is to remove those nodes with few connections, which process can be seen in Lei \& Lin (2022). Finally, we have two adjacency matrices $\cX, \cY\in\{0,1\}^{4722 \times 4722}$, each representing a network corresponding to 4722 genes.

\textbf{Result and Interpretation.}
The following results show the difference between the gene co-expression networks of E40 and E50 using the proposed method based on the maximum entry-wise deviation. Prior to using our method, we select the number of communities to be $K_{E40} = K_{E50} = 8$, which is considered reasonable, as described in Lei \& Lin (2022). Then, we use $T_n$ and $T_n^{boot}$ as test statistic, and obtain $T_n = 355748.3$ and $T_n^{boot} = 83.35$, respectively. Since, $Q_{0.95} = 3.41$ for the Gumbel distribution, we reject $H_0: (g_{E40}, B_{E40}) = (g_{E50}, B_{E50})$ at the level of 0.05. As the discussion in Lei \& Lin (2022), as the development occurs from 40 days to 50 days in the embryo, there are different gene communities that are most connected. From 40 days in embryo, community 1 was highly coordinated (i.e. densely connected), and to 40 days in embryo, community 3 was highly coordinated. Hence, it can be considered that the two networks in two periods E40 and E50 come from two different stochastic block models. The results are similar to that of Liu et al. (2018) and Lei \& Lin (2022).

\subsection{International trade data}

Now, we study an international trade dataset originally analyzed by Westveld \& Hoff (2011), containing yearly international trade data between $n = 58$ countries from 1981 to 2000. For this network, an adjacency matrix $A_t$ can be formed by first considering a weight matrix $W_t$ with $W_{ijt} = \mathrm{Trade}_{ijt} + \mathrm{Trade}_{jit}$ in given year $t$, where $\mathrm{Trade}_{ijt}$ denotes the value of exports from country $i$ to country $j$ in year $t$. Finally, we define $A_{ijt} = 1$ if $W_{ijt} \geq W_{0.5,t}$, and $A_{ijt} = 0$ otherwise, where $W_{0.5,t}$ denotes the 50th percentile of $\{W_{ijt}\}_{1\leq i < j\leq n}$ in year $t$. In this article, we focus on the international trade networks in 1995 and 1999. Thus, we can obtain two sample networks $\cX$ and $\cY$ in 1995 and 1999. For the network in 1995, the number of communities is estimated to be 7 in Hu et al. (2021). For the network in 1999, using the corrected Bayesian information criterion, the number of communities in 1999 is also estimated to be $7$. Then, we set $\widehat{K}_x = \widehat{K}_y = 7$ and continue to implement the proposed method. We obtain $T_n = 113.7352$ and $T_n^{boot} = 92.46558$. Since, $Q_{0.95} = 3.41$ for the Gumbel distribution, we reject $H_0: (g_{x}, B_{x}) = (g_{y}, B_{y})$ at the level of 0.05.  Hence, we can assert that the trade situation of different countries has changed in the four years from 1995 to 1999.

\section{DISSCUSSION}

\begin{figure}[h]
\centering
	\includegraphics[width = 8cm]{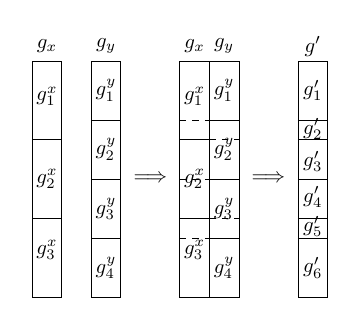}
	\caption{An example of a community combination}
	\label{figure:communities}
\end{figure}

In this article, we have proposed a novel two-sample test statistic based on the maximum entry-wise deviation of staggered centered and rescale observed adjacency matrix, and have demonstrated that the asymptotic null distribution of the test statistic is a Gumbel distribution when $K_x=K_y=o(n/\log^2n)$. This test has extended the method of Hu et al.(2021) to two samples. In our study, the difference between two networks is assessed by the sum of two residual matrices, where the centered and rescaled matrix of another sample is obtained using the estimation of the label and the probability matrix of one sample. Empirically, we have demonstrated that the size and the power of the test are valid. In this article, we assume that the numbers of communities for $\cX$ and $\cY$ are equal. In reality, we should determine whether $K_x$ is equal to $K_y$. In fact, we can independently estimate the number of communities by some existing methods, such as the sequentially testing methods (Lei, 2016: Hu et al., 2021) and methods based on the model selection (Hu et al., 2020). If $\widehat{K}_x=\widehat{K}_y$, the proposed method can be directly used. When $K_x \neq K_y$, we can consider the new community structure by combining two stochastic block models. Figure \ref{figure:communities} gives a diagram. Let $Z_x \in \{0,1\}^{n\times K_x}, Z_y \in \{0,1\}^{n\times K_y}, Z' \in \{0,1\}^{n\times K'}$ be membership matrices. Note that each row of $Z_x$ (or $Z_y$) has exactly one entry that is nonzero. For two nodes $i_1$ and $i_2$, $Z'_{i_1\cdot} = Z'_{i_2\cdot}$ if and only if $\bar{Z}_{i_{1}\cdot} = \bar{Z}_{i_{2}\cdot}$, where $\bar{Z} = (Z_x, Z_y) \in \{0, 1\}^{n\times(K_x+K_y)}$ be the combined membership matrix. Hence, we can use $\widehat{g}_x$ and $\widehat{g}_y$ to obtain $g'$, then obtain $\widehat{B}_x$ and $\widehat{B}_y$ by $g'$ and $K'$. Then, we can use our proposed method to test the difference between two samples after combining two community structures.

It is worth noting that the proposed testing method works well when the network is dense and $K$ is small in our simulation studies. However, when the network is sparse or $K$ is large, condition \eqref{eq:consistent} may be violated, i.e., the exact community label estimator is hard to be obtained. Hence, it would be of interest to investigate the possibility of the sparse network and big $K$ in future work. In addition, note that although the proposed method can identify whether the two models are the same through two samples, we cannot know whether this difference is caused by changes in $g$ or $B$, which is crucial in practical applications. Intuitively, we can judge whether $g_x$ is equal to $g_y$ by the strong consistent estimator $\widehat{g}_x$ and $\widehat{g}_y$. However, there is no theoretical guarantee. This issue will be considered in future work. In order to better improve the two-sample test of the stochastic block model, we will continue to study this issue in future work.

\section{APPENDIX}\label{Appendix}
We start with three lemmas that will be used in the proof. The following Poisson approximation result is essentially a special case of Theorem 1 in Arratia et al. (1989).

\begin{lemma}{lemma 1.}{[Arratia et al. (1989)]}
\label{lemma:Arratia}
Let $I$ be an index set and $\{\mathbf{B}_{\alpha},\alpha\in I\}$ be a set of subsets of $I$, that is, $\mathbf{B}_{\alpha}\subset I$. Let also $\{\eta_\alpha,\alpha\subset I\}$ be random variables. For a given $t\in\mathbb{R}$, set $\lambda = \sum_{\alpha\in I}\bbP\{\eta_\alpha>t\}$. Then
\[
\left|\bbP\{\max_{\alpha\in I}\eta_\alpha \leq t\}-e^{-\lambda}\right|\leq(1\wedge\lambda^{-1})(b_1+b_2+b_3),
\]
where
\[
b_1 = \sum_{\alpha\in I}\sum_{\beta\in\mathbf{B}_{\alpha}}\bbP\{\eta_\alpha>t\}\bbP\{\eta_\beta>t\}, b_2 = \sum_{\alpha\in I}\sum_{\alpha\neq\beta\in\mathbf{B}_{\alpha}}\bbP\{\eta_\alpha>t,\eta_\beta>t\},
\]
\[
b_3 = \sum_{\alpha\in I} \bbE|\bbP\{\eta_{\alpha}>t|\sigma(\eta_\beta,\beta\neq{\mathbf{B}_{\alpha})\}}-\bbP\{\eta_\beta>t\}|,
\]
and $\sigma(\eta_\beta,\beta\notin\mathbf{B}_{\alpha})$ is the $\sigma-$algebra generated by $\{\eta_\beta,\beta\notin\mathbf{B}_{\alpha}\}$. In particular, if $\eta_\alpha$ is independent of $\{\eta_\beta,\beta\notin\mathbf{B}_{\alpha}\}$ for each $\alpha$ then $b_3 = 0$.
\end{lemma}

\begin{lemma}{lemma 2.}{[Chen (1990)]}
\label{lemma:chen}
Suppose $\xi_1, \ldots,\xi_n$ are i.i.d random variables with $\bbE\xi_1=0$ and $\bbE\xi_1^2=1$. Set $S_n=\sum_{i=1}^n\xi_i$. Let $0<\alpha\leq 1$ and $\{a_n:n\geq 1\}$ satisfy that $a_n\rightarrow\infty$ and $a_n=o(n^{\alpha/(2(2-\alpha))})$. If $\bbE e^{t_0|\xi_1|^\alpha}<\infty$ for some $t_0>0$, then
$$\lim_n\frac{1}{a_n^2}\log \bbP\{\frac{S_n}{\sqrt{n}a_n}\geq \mu\}=-\frac{\mu^2}{2}$$
for any $\mu>0$.
\end{lemma}

\begin{lemma}{lemma 3.}{[Cai \& Jiang (2011)]}
\label{lemma:cai}
Suppose $\xi_1,\ldots,\xi_n$ are i.i.d random variables with $\bbE\xi_1=0$ and $\bbE\xi_1^2=1$ and $\bbE e^{t_0|\xi_1|^\alpha}<\infty$ for some $t_0 > 0$ and $0<\alpha\leq 1$. Set	$S_n=\sum_{i=1}^n\xi_i$ and $\beta = \alpha/(2+\alpha)$. Then, for any $\{p_n: n\geq 1\}$ with $0<p_n\rightarrow\infty$ and $\log p_n = o(n^{\beta})$ and $\{y_n: n\geq 1\}$ with $y_n\rightarrow y >0$,
\[
\bbP\{\dfrac{S_n}{\sqrt{n\log p_n}}>y_n\}\sim\dfrac{p_n^{-y_n^2/2}(\log p_n)^{-1/2}}{\sqrt{2\pi}y}
\]
as $n\rightarrow\infty$.
\end{lemma}

\subsection{Proof of Theorem 1}
Although the idea of proof of the Theorem 1 is similar to Theorem 1 in Hu et al. (2021), there are some differences in some details, such as decomposition for some quantities. Next, we prove Theorem 1.

Since, $H_0: (g_x, B_x) = (g_y, B_y)$, we denote $g = g_x = g_y$ and $B = B_x = B_y$ for simplicity. By Bernstein's inequality, we have
\[
\bbP\{|\widehat{B}_{uv}-B_{uv}|>\dfrac{\epsilon\log n}{\sqrt{n_un_v}}\}\leq2\exp\left(-\dfrac{\epsilon^2\log^2n}{2B_{uv}(1-B_{uv})+\frac{2}{3}\epsilon\log n/\sqrt{n_un_v}}\right)
\]
for any $\epsilon>0$, According to Assumption 2, we can get
\[
\widehat{B}_{uv}-B_{uv}=o_p(\dfrac{\log n}{\sqrt{n_un_v}})=o_p(\dfrac{K\log n}{n})
\]
uniformly in $u$ and $v$ as $n\rightarrow\infty$.

Notice that
\begin{align*}
	\tilde{\rho}_{iv,x} & := \sum\limits_{j\in g_y^{-1}(v)/\{i\}}\dfrac{X_{ij} - \widehat{B}^y_{g_y(i) g_y(j)}}{\sqrt{\widehat{B}^y_{g_y(i)g_y(j)}\left(1-\widehat{B}^y_{g_y(i)g_y(j)}\right)}}\\
	&\ = \sum\limits_{j\in g^{-1}(v)/\{i\}}\dfrac{X_{ij} - B_{g(i) g(j)} + o_p(\frac{K\log n}{n})}{\sqrt{B_{g(i)g(j)}\left(1-B_{g(i)g(j)}\right)}}\left(1+o_p(\sqrt{\dfrac{K\log n}{n}})\right)\\
	&\ = \rho_{iv,x} + \rho_{iv,x}o_p(\sqrt{\dfrac{K\log n}{n}}) + o_p(1),
\end{align*}
similarly,
\begin{align*}
	\tilde{\rho}_{iv,y} & := \sum\limits_{j\in g_x^{-1}(v)/\{i\}}\dfrac{Y_{ij} - \widehat{B}^x_{g_x(i)g_x(j)}}{\sqrt{\widehat{B}^x_{g_x(i)g_x(j)}\left(1-\widehat{B}^x_{g_x(i)g_x(j)}\right)}}\\
	&\ = \sum\limits_{j\in g^{-1}(v)/\{i\}}\dfrac{Y_{ij} - B_{g(i) g(j)} + o_p(\frac{K\log n}{n})}{\sqrt{B_{g(i)g(j)}\left(1-B_{g(i)g(j)}\right)}}\left(1+o_p(\sqrt{\dfrac{K\log n}{n}})\right)\\
	&\ = \rho_{iv,y} + \rho_{iv,y}o_p(\sqrt{\dfrac{K\log n}{n}}) + o_p(1),
\end{align*}
where $\rho_{iv,x} = \sum\limits_{j\in g^{-1}(v)/\{i\}}\dfrac{X_{ij} - B_{g(i) g(j)}}{\sqrt{B_{g(i)g(j)}\left(1-B_{g(i)g(j)}\right)}}$ and $\rho_{iv,y} = \sum\limits_{j\in g^{-1}(v)/\{i\}}\dfrac{Y_{ij} - B_{g(i) g(j)}}{\sqrt{B_{g(i)g(j)}\left(1-B_{g(i)g(j)}\right)}}$.

Let
\begin{align*}
	L_{n,0} & := \max_{i,v}|\tilde{\rho}_{iv}| \\
	&\ = \max_{i,v}\left| \dfrac{1}{\sqrt{\left|g_x^{-1}(v)/\{i\}\right|+\left|g_y^{-1}(v)/\{i\}\right|}}\left(\tilde{\rho}_{iv,x} + \tilde{\rho}_{iv,y}\right) \right|\\
	&\ = \max_{i,v}\left| \dfrac{1}{\sqrt{2\left|g^{-1}(v)/\{i\}\right|}}\left[\rho_{iv,x}+\rho_{iv,y} + (\rho_{iv,x}+\rho_{iv,y})o_p(\sqrt{\dfrac{K\log n}{n}}) + o_p(1)\right]\right| \\
	&\ = L_{n,1} + L_{n,1}o_p(\sqrt{\dfrac{K\log n}{n}}) + o_p(1),
\end{align*}
where
\begin{align*}
	L_{n,1} & := \max_{i,v}\left|\rho_{iv}\right| \\
	&\ = \max_{i,v}\left|\dfrac{1}{\sqrt{2\left|g^{-1}(v)/\{i\}\right|}}\left(\rho_{iv,x}+\rho_{iv,y}\right)\right|\\
	&\ = \max_{i,v}\left| \dfrac{1}{\sqrt{\left|g^{-1}(v)/\{i\}\right|}}\sum\limits_{j\in g^{-1}(v)/\{i\}}\dfrac{X_{ij} + Y_{ij} - 2B_{g(i) g(j)}}{\sqrt{2B_{g(i)g(j)}\left(1-B_{g(i)g(j)}\right)}} \right|.
\end{align*}

If $K=o(n/\log^2n)$ and $L_{n,1} = O_p(\sqrt{\log n})$, we have 
\[
L_{n,0} = L_{n,1}+o_p(1).
\]

Thus, to prove Theorem 1 \eqref{eq:gumbel}, it is sufficient to show
\[
\lim_{n}\bbP\left\{L_{n,1}^2-2\log(2Kn)+\log\log(2Kn)\leq y\right\}=\exp\left\{-\dfrac{1}{2\sqrt{\pi}}e^{-y/2}\right\}.
\]

Let $y_n=\sqrt{y+2\log(2Kn)-\log\log(2Kn)}, I=\{(i,v)|1\leq i\leq n,1\leq v\leq K\},\mathbf{B}_{iv}=\{(s,t)\in I/(i,v)|s=i\}$. Then $|\mathbf{B}_{iv}|=K-1$. Note that $\bbE\{\dfrac{X_{ij} + Y_{ij} - 2B_{g(i) g(j)}}{\sqrt{2B_{g(i)g(j)}\left(1-B_{g(i)g(j)}\right)}}\}=0$ and $\bbE\{\dfrac{X_{ij} + Y_{ij} - 2B_{g(i) g(j)}}{\sqrt{2B_{g(i)g(j)}\left(1-B_{g(i)g(j)}\right)}}\}^2=1$. By Lemma 1, we have
\[
\left|\bbP\{\max_{i,v}|\rho_{iv}|\leq y_n\}-e^{-\lambda_n} \right|\leq b_1 + b_2,
\]
where $\lambda_n = \sum_{i,v}\bbP\{|\rho_{iv}|>y_n\}$. By Lemma 3, we have
\begin{align*}
	\bbP\{\rho_{iv}>y_n\} & = \bbP\{\dfrac{\rho_{iv}}{\sqrt{\log(2Kn)}}>\sqrt{\dfrac{y+2\log(2Kn)-\log\log(2Kn)}{\log(2Kn)}}\} \\
	& \sim (2Kn)^{-\frac{y+2\log(2Kn)-\log\log(2Kn)}{\log(2Kn)}}(\log(2Kn))^{-1/2}/(2\sqrt{\pi})\\
	& = (2 K n)^{-1}(2 K n)^{-\frac{y}{2 \log (2 K n)}}(2 K n)^{\frac{\log \log (2 K n)}{2 \log (2 K n)}}(\log (2 K n))^{-\frac{1}{2}} /(2 \sqrt{\pi}) \\
	& = (2 K n)^{-1} e^{-\frac{y}{2 \log (2 K n)} \log (2 K n)} e^{\frac{\log \log (2 K n)}{2 \log (2 K n)} \log (2 K n)}(\log (2 K n))^{-\frac{1}{2}} /(2 \sqrt{\pi}) \\
	& = (2 K n)^{-1} e^{-y/2} e^{\log (\log (2 K n))^{\frac{1}{2}}}(\log (2 K n))^{-\frac{1}{2}} /(2 \sqrt{\pi}) \\
	& = (2 K n)^{-1} e^{-y/2}(\log (2 K n))^{\frac{1}{2}}(\log (2 K n))^{-\frac{1}{2}} /(2 \sqrt{\pi}) \\
	& = (2 K n)^{-1} e^{-y/2} /(2 \sqrt{\pi}).
\end{align*}
Hence,
\begin{align*}
	\lambda_n & = \sum_{i,v}\bbP\{|\rho_{iv}|>y_n\} \\
	& = Kn\dfrac{(Kn)^{-1}}{2\sqrt{\pi}}e^{-y/2} \\
	& = \dfrac{1}{2\sqrt{\pi}}e^{-y/2}.
\end{align*}

Meanwhile,
\begin{align*}
	b_1 & = \sum_{\alpha\in I}\sum_{\beta\in\mathbf{B}_{\alpha}}\bbP\{\eta_\alpha>y_n\}\bbP\{\eta_{\alpha}>y_n\} \\
	& < 4K^2ne^{-y-2\log(2Kn)+\log\log(2Kn)} \\
	& = e^{\log(4K^2n)-y-2\log(2Kn)+\log\log(2Kn)} \\
	& = o(1),
\end{align*}
\begin{align*}
	b_2 & = \sum_{\alpha\in I}\sum_{\alpha\neq\beta\in\mathbf{B}_{\alpha}}\bbP\{\eta_\alpha>y_n, \eta_{\alpha}>y_n\} \\
	& < 4K^2ne^{-y-2\log(2Kn)+\log\log(2Kn)} \\
	& = e^{\log(4K^2n)-y-2\log(2Kn)+\log\log(2Kn)} \\
	& = o(1).
\end{align*}

Thus, we have
\[
\lim_{n}\bbP\{L_{n,1}^2\leq y_n\} = \exp\left\{-\dfrac{1}{2\sqrt{\pi}}e^{-y/2}\right\},
\]
Combining this with $L_{n,0} = L_{n,1} + o_p(1)$, we know that \eqref{eq:gumbel} hlods.\hfill $\square$

\subsection{Proof of Theorem 2}

Although the idea of proof of the Theorem 2 is similar to Theorem 2 in Hu et al. (2021), there are some differences in some details, such as decomposition for some quantities. Next, we prove Theorem 2.

Similar to the proof of Theorem 1, we have
\[
\widehat{B}_{uv}^x-B_{uv}^x=o_p(\dfrac{K\log n}{n}),\ \mathrm{and}\ \widehat{B}_{uv}^y-B_{uv}^y=o_p(\dfrac{K\log n}{n}).
\]

Let 
\[
\rho_{iv,x} = \dfrac{1}{\sqrt{|g_y^{-1}(v)/\{i\}|}}\sum\limits_{j\in g_y^{-1}(v)/\{i\}}\dfrac{X_{ij} - B_{g_x(i)g_x(j)}^x}{\sqrt{B_{g_y(i)g_y(j)}^y\left(1-B_{g_y(i)g_y(j)}^y\right)}},
\]
\begin{align*}
	\tilde{\rho}_{iv,x} & := \dfrac{1}{\sqrt{|g_y^{-1}(v)/\{i\}|}}\sum\limits_{j\in g_y^{-1}(v)/\{i\}}\dfrac{X_{ij} - \widehat{B}^y_{g_y(i) g_y(j)}}{\sqrt{\widehat{B}^y_{g_y(i)g_y(j)}\left(1-\widehat{B}^y_{g_y(i)g_y(j)}\right)}}\\
	&\ = \dfrac{1}{\sqrt{|g_y^{-1}(v)/\{i\}|}} \\
	& \qquad \times\sum\limits_{j\in g_y^{-1}(v)/\{i\}}\dfrac{X_{ij} -B_{g_x(i)g_x(j)}^x + B_{g_x(i)g_x(j)}^x - B^y_{g_y(i) g_y(j)} + B^y_{g_y(i) g_y(j)} - \widehat{B}^y_{g_y(i) g_y(j)}}{\sqrt{\widehat{B}^y_{g_y(i)g_y(j)}\left(1-\widehat{B}^y_{g_y(i)g_y(j)}\right)}} \\
	&\ = \dfrac{1}{\sqrt{|g_y^{-1}(v)/\{i\}|}}\sum\limits_{j\in g_y^{-1}(v)/\{i\}}\dfrac{X_{ij} -B_{g_x(i)g_x(j)}^x + B_{g_x(i)g_x(j)}^x - B^y_{g_y(i) g_y(j)} + o_p(\dfrac{K\log n}{n})}{\sqrt{B^y_{g_y(i)g_y(j)}\left(1-B^y_{g_y(i)g_y(j)}\right)}}\\
	&\qquad \times \left(1+o_p(\sqrt{\dfrac{K\log n}{n}})\right) \\
	&\ = \rho_{iv,x} + \dfrac{1}{\sqrt{|g_y^{-1}(v)/\{i\}|}}\sum\limits_{j\in g_y^{-1}(v)/\{i\}}\dfrac{B_{g_x(i)g_x(j)}^x - B^y_{g_y(i) g_y(j)}}{\sqrt{B^y_{g_y(i)g_y(j)}\left(1-B^y_{g_y(i)g_y(j)}\right)}} + o_p(1).
\end{align*}

By Hoeffding's inequality, we have 
\begin{align*}
	& \bbP\{\max_{i,v}|\dfrac{1}{\sqrt{|g_y^{-1}(v)/\{i\}|}}\sum\limits_{j\in g_y^{-1}(v)/\{i\}}\dfrac{X_{ij} - B_{g_x(i)g_x(j)}^x}{\sqrt{B_{g_y(i)g_y(j)}^y\left(1-B_{g_y(i)g_y(j)}^y\right)}}|>t\} \\
	\leq &\ \sum_{i,v}\bbP\{|\sum\limits_{j\in g_y^{-1}(v)/\{i\}}\dfrac{X_{ij} - B_{g_x(i)g_x(j)}^x}{\sqrt{B_{g_y(i)g_y(j)}^y\left(1-B_{g_y(i)g_y(j)}^y\right)}}|>t\sqrt{|g_y^{-1}(v)/\{i\}|}\} \\
	\leq &\ 2e^{\log(Kn)-2C_1^2t^2}.
\end{align*}
Hence, $\max_{i,v}|\rho_{iv,x}| = O_p(\sqrt{\log n})$. Denote $$\dfrac{1}{\sqrt{|g_y^{-1}(v)/\{i\}|}}\sum\limits_{j\in g_y^{-1}(v)/\{i\}}\dfrac{B_{g_x(i)g_x(j)}^x - B^y_{g_y(i) g_y(j)}}{\sqrt{B^y_{g_y(i)g_y(j)}\left(1-B^y_{g_y(i)g_y(j)}\right)}}$$ by $\ell_{iv}(y,x)$, we have
\[
\max_{i,v}|\tilde{\rho}_{iv,x}| = \max_{i,v}|\ell_{iv}(y,x)| + O_p(\sqrt{\log n}).
\]

Similarly, we have
\[
\rho_{iv,y} = \dfrac{1}{\sqrt{|g_x^{-1}(v)/\{i\}|}}\sum\limits_{j\in g_x^{-1}(v)/\{i\}}\dfrac{Y_{ij} - B_{g_y(i)g_y(j)}^y}{\sqrt{B_{g_x(i)g_x(j)}^x\left(1-B_{g_x(i)g_x(j)}^x\right)}},
\]

\begin{align*}
	\tilde{\rho}_{iv,y} & := \dfrac{1}{\sqrt{|g_x^{-1}(v)/\{i\}|}}\sum\limits_{j\in g_x^{-1}(v)/\{i\}}\dfrac{Y_{ij} - \widehat{B}^x_{g_x(i) g_x(j)}}{\sqrt{\widehat{B}^x_{g_x(i)g_x(j)}\left(1-\widehat{B}^x_{g_x(i)g_x(j)}\right)}}\\
	&\ = \rho_{iv,y} + \ell_{iv}(x,y) + o_p(1).
\end{align*}
where $\ell_{iv}(x,y) = \dfrac{1}{\sqrt{|g_x^{-1}(v)/\{i\}|}}\sum\limits_{j\in g_x^{-1}(v)/\{i\}}\dfrac{B_{g_y(i)g_y(j)}^y - B^x_{g_x(i) g_x(j)}}{\sqrt{B^x_{g_x(i)g_x(j)}\left(1-B^x_{g_x(i)g_x(j)}\right)}}$, and
\[
\max_{i,v}|\tilde{\rho}_{iv,y}| = \max_{i,v}|\ell_{iv}(x,y)| + O_p(\sqrt{\log n}).
\]

Hence, 
\begin{align*}
	L_{n,0} & := \max_{i,v}|\tilde{\rho}_{iv}| \\
	&\ = \max_{i,v}|\dfrac{1}{\sqrt{\left|g_x^{-1}(v)/\{i\}\right|+\left|g_y^{-1}(v)/\{i\}\right|}}(\sqrt{\left|g_y^{-1}(v)/\{i\}\right|}\tilde{\rho}_{iv,x}+\sqrt{\left|g_x^{-1}(v)/\{i\}\right|}\tilde{\rho}_{iv,y})| \\
	&\ \geq \max_{i,v}|\dfrac{\sqrt{\left|g_x^{-1}(v)/\{i\}\right|}+\sqrt{\left|g_y^{-1}(v)/\{i\}\right|}}{\sqrt{\left|g_x^{-1}(v)/\{i\}\right| + \left|g_y^{-1}(v)/\{i\}\right|}} (\tilde{\rho}_{iv,x}\wedge\tilde{\rho}_{iv,y})| \\
	&\ = \dfrac{\sqrt{\left|g_x^{-1}(v)/\{i\}\right|}+\sqrt{\left|g_y^{-1}(v)/\{i\}\right|}}{\sqrt{\left|g_x^{-1}(v)/\{i\}\right| + \left|g_y^{-1}(v)/\{i\}\right|}} \max_{i,v}|(\tilde{\rho}_{iv,x}\wedge\tilde{\rho}_{iv,y})|
\end{align*}
By Assumptions 2 and 3, we have $L_{n,0}^2/\log n\rightarrow\infty$. 

Let $T_n=L_n^2(g_x, g_y)-2\log(2Kn)+\log\log(2Kn)$, as $n\rightarrow\infty$, we have 
\[
T_n/\log n\rightarrow\infty.
\]
\hfill$\square$

\subsection{The Bootstrap corrected test statistic under the degree-corrected stochastic block model}

For two sample networks $\cX$ and $\cY$, under the framework of the degree-corrected stochastic block model, the bootstrap corrected test statistic is calculated as the following:

1. Estimating $(\widehat{g}_x,\widehat{B}_x)$ and $(\widehat{g}_y,\widehat{B}_y)$ by the consistent clustering method and $\widehat{\theta}_x$ and $\widehat{\theta}_y$ using their maximum likelihood estimators. Calculate the statistic $T_n$ using $\cX, \cY, (\widehat{g}_x,\widehat{B}_x, \widehat{\theta}_x)$ and $(\widehat{g}_y,\widehat{B}_y, \widehat{\theta}_y)$;

2. For $m = 1,\ldots,M$, generate $\cX^{(m)}$ and $\cY^{(m)}$ from the edge-probability matrix $\widehat{P}_{ij} = (\widehat{P}_{ij}^x + \widehat{P}_{ij}^y)/2 = (\widehat{\theta}_i^x\widehat{\theta}_j^x\widehat{B}_{\widehat{g}_x(i)\widehat{g}_x(j)}^x + \widehat{\theta}_i^y\widehat{\theta}_j^y\widehat{B}_{\widehat{g}_y(i)\widehat{g}_y(j)}^y)/2$, and calculate $T_n^{(m)}$ based on $\cX^{(m)}$, $\cY^{(m)}$;

3. Using $\{T_n^{(m)}: m=1,\ldots,M\}$ to estimate the location and scale parameters $\widehat{\mu}$ and $\widehat{\beta}$ of the Gumbel distribution through maximum likelihood method.

4. The bootstrap corrected test statistic is calculated as
\[
T_{n}^{boot} = \mu + \beta\left(\dfrac{T_n - \widehat{\mu}}{\widehat{\beta}}\right),
\]
where $\mu = -2\log(2\sqrt{\pi})$ and $\beta = 2$.


\begin{thebibliography}{}
\bibitem{Airoldi:2008}
Airoldi, E. M., Blei, D. M., Fienberg, S. E., \& Xing, E. P. (2008). Mixed membership stochastic blockmodels. {\it Journal of machine learning research}, 9:1981 - 2014.


\bibitem{Amini:2013}
Amini, A. A., Chen, A., Bickel, P. J., \& Levina, E. (2013). Pseudo-Likelihood Methods for Community Detection in Large Sparse Networks. {\it The Annals of Statistics}, 41:2097 - 2122.

\bibitem{Arratia:1989}
Arratia, R., Goldstein, L., \& Gordon, L. (1989). Two moments suffice for Poisson approximations: The Chen-Stein method. {\it The Annals of Probability}, 17:9 - 25.

\bibitem{Bakken:2016}
Bakken, T. E., Miller, J. A., Ding, S.-L., Sunkin, S. M., Smith, K. A., Ng, L., Szafer, A., Dalley, R. A., Royall, J. J., Lemon, T., et al (2016). A comprehensive transcriptional map of primate brain development. {\it Nature}, 535:367 - 375.

\bibitem{Barnett:2016}
Barnett, I. \& Onnela, J.-P. (2016). Change point detection in correlation networks. {\it Scientic reports}, 6:18893.

\bibitem{Bickel:2009}
Bickel, P. J. \& Chen, A. (2009). A nonparametric view of network models and Newman-Girvan and other modularities. {\it Proceedings of the National Academy of Sciences}, 106:21068 - 21073.

\bibitem{Bickel:2015}
Bickel, P. J., \& Sarkar, P. (2015), Hypothesis testing for automated community detection in networks, {\it Journal of the Royal Statistical Society, Series B}, 78:253–273.

\bibitem{Cai:2011}
Cai, T. T. \& Jiang, T. (2011). Limiting laws of coherence of random matrices with applications to testing covariance structure and construction of compressed sensing matrices. {\it The Annals of Statistics}, 39:1496 - 1525.

\bibitem{Chen:1990}
Chen, X. (1990). Probabilities of moderate deviations for $B$-valued independent random vectors. {\it Chinese Annals of Mathematics}, 11:621 - 629.

\bibitem{Chen:2021a}
Chen, L., Zhou, J. \& Lin, L. (2021a). Hypothesis testing for populations of networks. {\it Communications in Statistics - Theory and Methods}, online.

\bibitem{Chen:2021b}
Chen, L., Josephs, N., Lin, L., Zhou, J. \& Kolaczyk, E. D. (2021b). A spectral-based framework for hypothesis testing in populations of networks. {\it Statistica Sinica}, online.

\bibitem{Choi:2012}
Choi, D. S., Wolfe, P. J., \& Airoldi, E. M. (2012). Stochastic blockmodels with a growing number of classes. {\it Biometrika}, 99:273 - 284.

\bibitem{Daudin:2008}
Daudin, J.-J., Picard, F., \& Robin, S. (2008). A mixture model for random graphs. {\it Statistics and Computing}, 18:173 - 183.

\bibitem{Gao:2017}
Gao, C., Ma, Z., Zhang, A. Y., \& Zhou, H. H. (2017). Achieving optimal misclassification proportion in stochastic blockModels. {\it The Journal of Machine Learning Research}, 18:1980 - 2024.

\bibitem{Ghoshdastidar:2018}
Ghoshdastidar, D., \& von Luxburg, U. (2018). Practical methods for graph two-sample testing. In Advances in Neural Information Processing Systems, 3019-3028. Montr\'{e}al, Canada.

\bibitem{Ghoshdastidar:2020}
Ghoshdastidar, D., Gutzeit, M., Carpentier, A. \& Von Luxburg, U.(2020). Two-sample hypothesis testing for inhomogeneous random graphs. {\it The Annals of Statistics}, 48: 2208 - 2229.

\bibitem{Holland:1983}
Holland, P. W., Laskey, K. B., \& Leinhardt, S. (1983). Stochastic blockmodels: First steps. {\it Social Networks}, 5:109 - 137.

\bibitem{Hu:2020}
Hu, J., Qin, H., Yan, T., , \& Zhao, Y. (2020). Corrected Bayesian Information Criterion for Stochastic Block Models. {\it Journal of the American Statistical Association}, 115:1771 - 1783.

\bibitem{Hu:2021}
Hu, J., Zhang, J., Qin, H., Yan, T., \& Zhu, J. (2021). Using Maximum Entry-Wise Deviation to Test the Goodness of Fit for Stochastic Block Models. {\it Journal of The American Statistical Association}, 116:1373 - 1382.

\bibitem{Jin:2015}
Jin, J. (2015). Fast Community Detection by SCORE. {\it The Annals of Statistics}, 43:57- 89.

\bibitem{Karrer:2011}
Karrer, B. \& Newman, M. E. J. (2011). Stochastic blockmodels and community structure in networks. {\it Physical Review. E}, 83:016107.

\bibitem{Langfelder:2008}
Langfelder, P. \& Horvath, S. (2008). WGCNA: An R package for weighted correlation network analysis. {\it BMC Bioinformatics}, 9:559.

\bibitem{Le:2022}
Le, C. M. \& Levina, E. (2022). Estimating the number of communities in networks by spectral methods. {\it Electronic Journal of Statistics}, 16:3315 - 3342.

\bibitem{Lei:2006}
Lei, J. (2016). A goodness-of-fit test for stochastic block models. {\it The Annals of Statistics}, 44:401 - 424.

\bibitem{Lei:2022}
Lei, J. \& Lin, K. Z. (2022). Bias-adjusted spectral clustering in multi-layer stochastic block models. {\it Journal of the American Statistical Association}, online.

\bibitem{Lei:2015}
Lei, J. \& Rinaldo, A. (2015). Consistency of spectral clustering in stochastic block models. {\it The Annals of Statistics}, 43:215 - 237.

\bibitem{Liu:2018}
Liu, F., Choi, D., Xie, L., \& Roeder, K. (2018). Global spectral clustering in dynamic networks. {\it Proceedings of the National Academy of Sciences}, 115:927 - 932.

\bibitem{Newman:2006}
Newman, M. E. J. (2006). Modularity and community structure in networks. {\it Proceedings of the National Academy of Sciences of the United States of America}, 103:8577 - 8582.

\bibitem{Newman:2004}
Newman, M. E. J. \& Girvan, M. (2004). Finding and evaluating community structure in networks. {\it Physical review. E}, 69:026113.

\bibitem{Nowicki:2001}
Nowicki, K. \& Snijders, T. (2001). Estimation and prediction for stochastic block structures. {\it Journal of The American Statistical Association}, 96:1077 - 1087.

\bibitem{Rohe:2011}
Rohe, K., Chatterjee, S., \& Yu, B. (2011). Spectral clustering and the high-dimensional stochastic blockmodel. {\it The Annals of Statistics}, 39:1878 - 1915.

\bibitem{Rho:2014}
Rohe, K., Qin, T., \& Fan, H. (2014). The highest dimensional stochastic blockmodel with a regularized estimator. {\it Statistica Sinica}, 24:1771 - 1786.

\bibitem{Saldna:2017}
Sald{\~n}a, D. F., Yu, Y., \& Feng, Y. (2017). How Many Communities Are There? {\it Journal of Computational and Graphical Statistics}, 26:171 - 181.

\bibitem{Sarkar:2015}
Sarkar, P. \& Bickel, P. J. (2015). Role of normalization in spectral clustering for stochastic blockmodels. {\it The Annals of Statistics}, 43:962 - 990.

\bibitem{Snijders:1997}
Snijders, T. \& Nowicki, K. (1997). Estimation and prediction for stochastic block structures for graphs with latent block structure. {\it Journal of Classification}, 14:75 - 100.

\bibitem{Steinhaeuser:2010}
Steinhaeuser, K. \& Chawla, N. V. (2010). Identifying and evaluating community structure in complex networks. {\it Pattern Recognition Letters}, 31:413 - 421.

\bibitem{Tang:2017}
Tang, M., Athreya, A., Sussman, A. L., Lyzinski, V. \& Priebe, C. E. (2017). A nonparametric two-sample hypothesis testing problem for random graphs. {\it Bernoulli}, 23:1599 - 1630.


\bibitem{Wang:2021}
Wang, J., Zhang, J., Liu, B., Zhu, J., \& Guo, J. (2021). Fast network community detection with profile-pseudo likelihood methods. {\it Journal of the American Statistical Association}, online.

\bibitem{Wang:2017}
Wang, Y. R. \& Bickel, P. J. (2017). Likelihood-based model selection for stochastic block models. {\it The Annals of Statistics}, 45:500 - 528.

\bibitem{Westveld:2011}
Westveld, A. H. \& Hoff, P. D. (2011). A Mixed effects model for longitudinal relational and network data with applications to international trade and conflict. {\it The Annals of Applied Statistics}, 5:843 - 872.

\bibitem{Wu:2022}
Wu, F., Kong, X. \& Xu, C. (2022). Test on stochastic block model: local smoothing and extreme value theory. {\it Journal of Systems Science and Complexity}, 35:1535 - 1556.

\bibitem{Zhang:2016}
Zhang, A. Y. \& Zhou, H. H. (2016). Minimax rates of community detection in stochastic block models. {\it The Annals of Statistics}, 44:2252 - 2280.

\bibitem{Zhao:2011}
Zhao, Y., Levina, E., \& Zhu, J. (2011). Community extraction for social networks. {\it Proceedings of the National Academy of Sciences of the United States of America}, 108:7321 - 7326.

\bibitem{Zhao:2012}
Zhao, Y., Levina, E., \& Zhu, J. (2012). Consistency of community detection in networks under degree-corrected stochastic block models. {\it The Annals of Statistics}, 40:2266 - 2292.

\bibitem{Zhou:2007}
Zhou, W. (2007). Asymptotic distribution of the largest off-diagonal entry of correlation matrices. {\it Transactions of the American Mathematical Society}, 359:5345 - 5363.



\end{thebibliography}
\end{document}